\documentclass[conference]{IEEEtran}
\IEEEoverridecommandlockouts
\usepackage{cite}
\usepackage{amsmath,amssymb,amsfonts}
\usepackage{algorithmic}
\usepackage{graphicx}
\usepackage{textcomp}
\usepackage{booktabs}
\usepackage{multirow}
\usepackage{xcolor}
\usepackage{comment}
\usepackage[activate]{pdfcprot}
\usepackage{todonotes}
\def\BibTeX{{\rm B\kern-.05em{\sc i\kern-.025em b}\kern-.08em
    T\kern-.1667em\lower.7ex\hbox{E}\kern-.125emX}}
\begin{document}

\newcommand{\nix}[1]{}

\newcommand{\PE}{\emph{PE}}
\newcommand{\PEs}{\emph{PEs}}
\newcommand{\HD}{H\!D}
\newcommand{\PM}{P\!M}

\title{SystemC Model of Power Side-Channel Attacks Against AI Accelerators: Superstition or not?
\thanks{This work has been funded by the German Ministry of Education and Research (BMBF) via the project VE-Jupiter (16ME0234).}
}

\author{\IEEEauthorblockN{Andrija Ne\v{s}kovi\'{c}\textsuperscript{1}, Saleh Mulhem\textsuperscript{1}, Alexander Treff\textsuperscript{2}, Rainer Buchty\textsuperscript{1}, Thomas Eisenbarth\textsuperscript{2}, and Mladen Berekovic\textsuperscript{1}}
\IEEEauthorblockA{\textit{\textsuperscript{1}Institute of Computer Engineering and \textsuperscript{2}Institute for IT Security} \\
\textit{University of Lübeck}\\
Lübeck, Germany \\
\{andrija.neskovic, saleh.mulhem, a.treff, rainer.buchty, thomas.eisenbarth, mladen.berekovic\}@uni-luebeck.de}
}

\maketitle
\begin{abstract}
As training artificial intelligence (AI) models is a lengthy and hence costly process, leakage of such a model's internal parameters is highly undesirable. In the case of AI accelerators, side-channel information leakage opens up the threat scenario of extracting the internal secrets of pre-trained models. Therefore, sufficiently elaborate methods for design verification as well as fault and security evaluation at the electronic system level are in demand.

In this paper, we propose estimating information leakage from the early design steps of AI accelerators to aid in a more robust architectural design. We first introduce the threat scenario before diving into SystemC as a standard method for early design evaluation and how this can be applied to threat modeling. We present two successful side-channel attack methods executed via SystemC-based power modeling: correlation power analysis and template attack, both leading to total information leakage. The presented models are verified against an industry-standard netlist-level power estimation to prove general feasibility and determine accuracy. Consequently, we explore the impact of additive noise in our simulation to establish indicators for early threat evaluation. The presented approach is again validated via a model-vs-netlist comparison, showing high accuracy of the achieved results. This work hence is a solid step towards fast attack deployment and, subsequently, the design of attack-resilient AI accelerators.
\end{abstract}


\section{Introduction}
The use of Artificial Intelligence (AI) is rapidly growing across all emerging technologies. One of the most important aspects is accelerating the AI inference process and building according hardware accelerators. An accelerator design's fault-tolerance mechanisms and other safety features are usually evaluated in the pre-silicon phase, whereas evaluation of the accelerator's physical security is performed in the post-silicon phase. Side channel attacks, especially power attacks, are considered a serious security threat leading to a vulnerable AI hardware component\nix{as an integrated circuit (IC)}. Recently, studies in the domain of AI accelerator design show that side-channel leakages of AI accelerators can be exploited to reveal industrial secrets such as the AI model architecture and its parameters~\cite{CSI_NN19,chabanne21NN,10.1145/3474376.3487284}. For instance, power attacks are deployed successfully to reverse engineer the AI model~\cite{CSI_NN19,KotaYOSHIDA20212020CIP0024}. Power attacks lead to copying the AI model and then distributing it as counterfeit intellectual property (IP). Therefore, a huge need exists for evaluating an AI accelerator's side-channel attack resistance in the early design steps (EDS) of integrated circuit (IC) design, such as \nix{IC design }security evaluation at register-transfer level (RTL), using gate-level netlists, or even earlier. Evaluation and investigation of security issues in EDS provide insight into the robustness of the later fabricated IC. Thus, design decisions made at high abstraction levels have significant impact on the whole design process.

\begin{figure}[t]
\includegraphics[width=1\hsize, trim={1cm 2.2cm 1.1cm 1cm}]{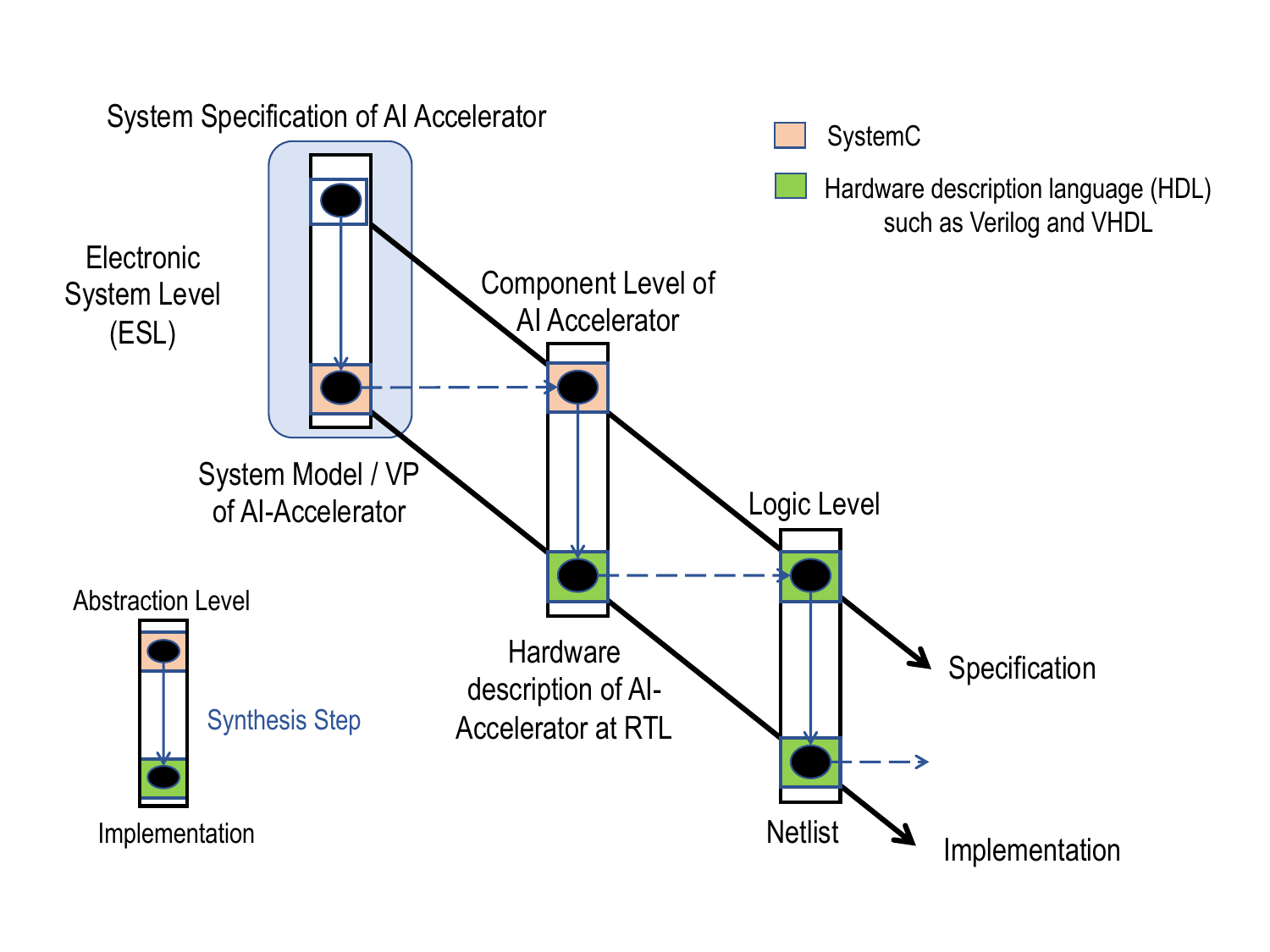}
\centering
\caption{AI Accelerator IC Design Process (Adapted from \cite{5247153}).}
\label{fig:HDProcess}
\end{figure}

Tools and platforms considering security evaluation in EDS were developed mainly to detect hardware Trojan circuitry \nix{inserted into ICs}~\cite{6653605} or to check the security rule in ICs~\cite{7543447}. Recent work confirms the importance of security evaluation in the EDS by demonstrating static and dynamic information flow analysis using Virtual Prototypes (VPs) \cite{Drechsler-Security-Paper, Drechsler-security-dynamic}. Simulators targeting side-channel evaluation have been studied in~\cite{SCA-simulators}. However, the reviewed tools consider only the software implementation of cryptographic algorithms executed on general purpose hardware (i.e. microcontrollers). \textit{The evaluation of side-channel attacks (SCA) and its impact on ICs are still an open issue in EDS:} To the best of our knowledge, no simulator exists targeting SCA evaluation of dedicated AI hardware accelerators. The missing SCA evaluation in EDS of IC is required to ensure the IC's dependability, together with existing reliability and safety tests~\cite{9869587}. 

Our work demonstrates previously shown power SCA by utilizing a dedicated power estimation model with the goal to evaluate the worst-case resilience of an AI accelerator's design at electronic system level (ESL). This approach was positively verified by a comparison of the SystemC model's behavior with a technology-synthesized gate-level netlist.

\subsection{Why a SystemC Model?}
SystemC is a solid candidate~\cite{9129426} for performing security evaluation in EDS, as it is one of the industry standards for hardware/software modelling at high abstraction levels.
Particularly, SystemC is C++-based and was originally conceived for hardware/software co-design, simulation, and functional verification~\cite{DrechslerBook2004}.
Over time, new design aspects such as fault evaluation~\cite{7880368} and power modeling~\cite{2852345, SystemC-VP-Security} were also addressed using SystemC.
The security assessment of IC design has recently received more attention \cite{7543447}, especially by SystemC in EDS~\cite{9129426}.
With proper power estimation models, SystemC can be utilized to simulate power attacks against ICs even at the ESL.

In order to clarify how to deploy SystemC in EDS, Fig.~\ref{fig:HDProcess} shows the top-down hardware design process of AI accelerators (a modified version of the \textit{Double roof model} described in~\cite{5247153,DRTeich2000}).
Starting from ESL, the requirements towards AI accelerators are specified and synthesized into a system model (most likely represented by a VP~\cite{6176558}).
The requirements for lower abstraction levels are derived based on this specification and implementation. At every abstraction level, the specification is transformed into an implementation with a synthesis step.
This work presents a SystemC model of an AI accelerator at ESL.


\subsection{Paper Contribution}
In this paper, we show how to evaluate power attacks against AI accelerators in EDS. 
For this, we build a SystemC model of a systolic-array-based AI accelerator hardware at ESL.
Using SystemC, the activation count of components can be annotated with a power-consumption model to generate power traces covering the hardware of AI accelerators.
We demonstrate a correlation power analysis (CPA) and template attacks (TA) based on our SystemC model in ideal conditions and explore the limits of these attacks in a noisy environment. Finally, we show the comparison of our model-based traces against power traces from a state of the art netlist level simulation to demonstrate the feasibility of the proposed model.

\section{Related Work}
The target of this work is to evaluate the benefits of using SystemC models to analyze side-channel information leakage at ESL in EDS.
Therefore, power side-channel attacks against AI accelerators are briefly discussed in this section, as well as modeling approaches utilizing SystemC in different areas.


\subsection{Power Attacks vs. AI Hardware Accelerator}
Several power attack scenarios against AI hardware accelerators have been proposed~\cite{9300276,KotaYOSHIDA20212020CIP0024}.
Power attacks exploit power consumption leakage from an accelerator executing a pre-trained AI model (simply AI model) to reveal its internal secrets.
In particular, an attacker uses an evaluation board attached to a targeted device~\cite{CPA-39555-5_29}, i.e., the AI Hardware accelerator, and captures power consumption traces of some data input. The attacker applies statistical analysis, e.g., Simple Power Analysis (SPA), Differential Power Analysis (DPA), or Correlation Power Analysis (CPA), on the data input and the power traces to recover the internal secrets of the AI model. For instance, DPA was deployed to extract the AI's secret parameters in \cite{9300276}. In \cite{KotaYOSHIDA20212020CIP0024}, CPA was applied against the systolic-array-based hardware accelerator of Deep Neural Networks (DNNs).

\subsection{Power Consumption Modeling Challenges} 
The power consumption of any CMOS computing platform includes two types: static (leakage) power consumption \(P_{static}\) and dynamic (switching) power consumption \(P_{dynamic}\) \cite{harris2010digital}. The total power for a computing platform can be modeled by:   
\begin{equation}
    P_{total} = P_{dynamic} + P_{static} 
    \label{eq:TOTALpow}
\end{equation}
\(P_{static}\) is the product of leakage current and the supply voltage \cite{JACOB2008847}, and \(P_{dynamic}\) indicates and quantifies transistor switching. Thus, \(P_{dynamic}\) provides a distinctive current profile. Therefore, power attacks mainly rely on \(P_{dynamic}\), which is considered the Achilles heel of any CMOS computing platforms \cite{soares2021hardware}. 

SystemC can be utilized for the activation count of hardware components at different abstraction levels. It plays a crucial role in simulating power attacks in EDS. Using SystemC to estimate a computing platform's power consumption poses several challenges: SystemC was used in \cite{6951886} to estimate the power consumption of different processor configurations based on pre-computed power values of its components, such as memories, register files, function units, etc. The proposed power model exhibited a 15\% prediction error. In \cite{2852345}, a black-box power model was introduced for digital signal processors (DSPs) in SystemC. The proposed power model does not require detailed insight into the individual components of the probed computing platform. The black-box power model exhibited a prediction error of less than 4\%. This prediction error is caused by the lack of information about power dissipation \(P_{static}\) of the targeted manufacturing technology. Therefore, the power consumption of a computing platform is rather difficult to model in SystemC. However, our work introduces a SystemC model that considers only dynamic power consumption \(P_{dynamic}\) to enable simulating power attacks.

\subsection{SystemC for Security Evaluation}
Utilizing a SystemC VP to evaluate security-critical systems on chips has already been demonstrated in~\cite{SystemC-VP-Security}. Beyond that, SystemC was proven successful for power-attack evaluation of cryptographic applications in~\cite{9129426}, such as RSA-based public-key cryptosystems and elliptic-curve cryptography.
This approach in~\cite{9129426} solely relies on a dedicated dynamic power consumption model, the so-called input-dependent model.
This model covers the arithmetic operations by assuming that there is no difference between simulated hardware or C++ operators.
The input-dependent model covers bit-shifts and comparisons as well, but lacks power modeling of registers, multiplexers, and other hardware components.
In order to extend SystemC power attacks analysis beyond cryptographic applications, this paper introduces power estimation models to cover systolic arrays.
By utilizing these power models, CPA and power-template attacks (TA) against AI accelerators are simulated.         

In the following sections, we first build a system model of a systolic-array-based AI accelerator and extend and modify power-consumption models proposed in~\cite{9129426, SystemC-VP-Security} to also cover additional components present in an AI hardware accelerator.
Then, we perform the proposed attacks. Finally, we verify our power estimation model and the validity of the attacked performed hereon against a state-of-the-art netlist-level power estimation tool.
\section{AI Accelerator ESL Model}

\begin{figure}[t]
\includegraphics[width=1\hsize, trim={0.5cm 0.8cm 0cm 0.5cm}]{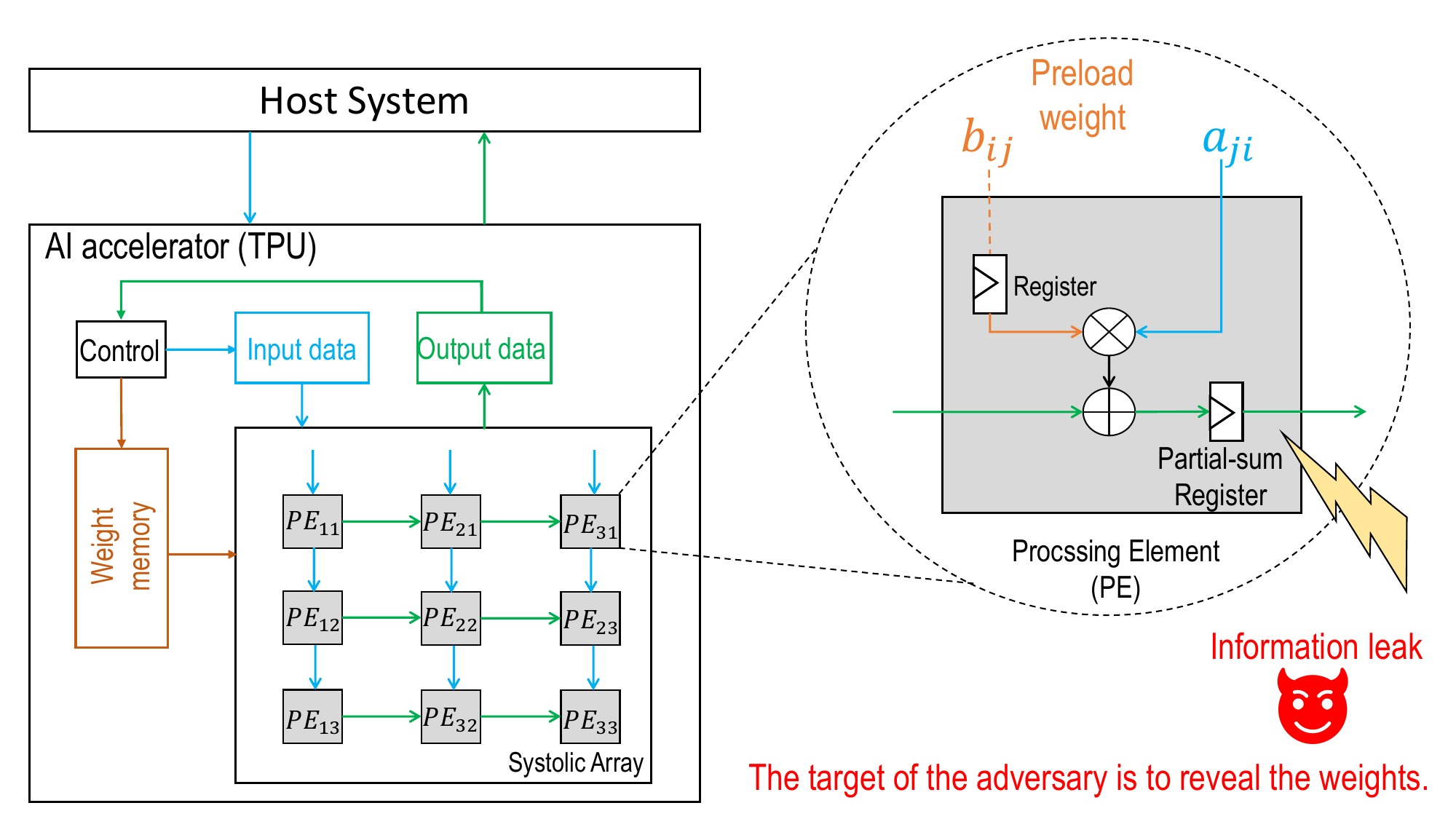}
\centering
\caption{Block diagram of TPU including the threat model.}
\label{fig:blockDiagram}
\end{figure}

To perform a security evaluation in EDS, an ESL model of an AI accelerator is required. For our approach, we use SystemC as the modelling language and start with a loosely-timed SystemC model 
of a systolic array for simulating the behaviour. By annotating this model with input-dependent power estimation capabilities, all components for SCA simulation can be provided.

\subsection{Systolic Array for Acceleration}
The inference process of AI applications requires frequent data access.
Such data-read operations from memory are very costly and time-consuming and
therefore should be avoided on edge accelerators in order to minimize power consumption and maximize performance. This can be addressed by using so-called systolic array architectures, featuring a number of benefits~\cite{1653825}.
Instead of accessing memory after every arithmetic operation, the systolic design approach utilizes multiple processing elements (\PEs) to avoid frequent memory access.
Each \PE\ performs a multiply-accumulate operation (MAC) as shown in Fig.~\ref{fig:blockDiagram}.
The partial result of an \PE\ is directly passed to another \PE\ without memory access. 
The realization of an array of \PEs\ in hardware can accelerate matrix multiplication, which is essential for accelerating the desired AI algorithm.

The matrix multiplication of \(A=(a_{ij})_{3\times 3}\) and \(B=(b_{ij})_{3\times 3}\) results in a matrix \(C=(c_{ij})_{3\times 3}\).
A systolic array accelerates such a matrix multiplication, where, a resulting element (\(c_{11}\)) is, for instance, calculated sequentially over \(3\) clock-cycles by performing 3 MACs in 3 different \PEs\ as follows \cite{KotaYOSHIDA20212020CIP0024}:
\begin{equation}
\label{eq:PEsInParallel}
\begin{array}{rlr}
Reg_{11} & =a_{11} \times b_{11}+0 & (t=1) \\
Reg_{21} & =a_{12} \times b_{21}+Reg_{11} & \\
& =a_{12} \times b_{21}+a_{11} \times b_{11} & (t=2) \\
Reg_{31} & =a_{13} \times b_{31}+Reg_{21} & \\
& =a_{13} \times b_{31}+a_{12} \times b_{21}+a_{11} \times b_{11} & (t=3)
\end{array}
\end{equation}
Where \(Reg_{ij}\) is a partial sum register of \(\PE_{ij}\) as shown in Fig.~\ref{fig:blockDiagram}. In our model, the weights and inputs are represented as 8\,bit integers, and the partial-sum results as 18\,bit integers.

\subsection{SystemC Model of the Accelerator}
\begin{figure}[t]
    \includegraphics[width=0.777\hsize, trim={2cm 1.5cm 6cm 0cm}]{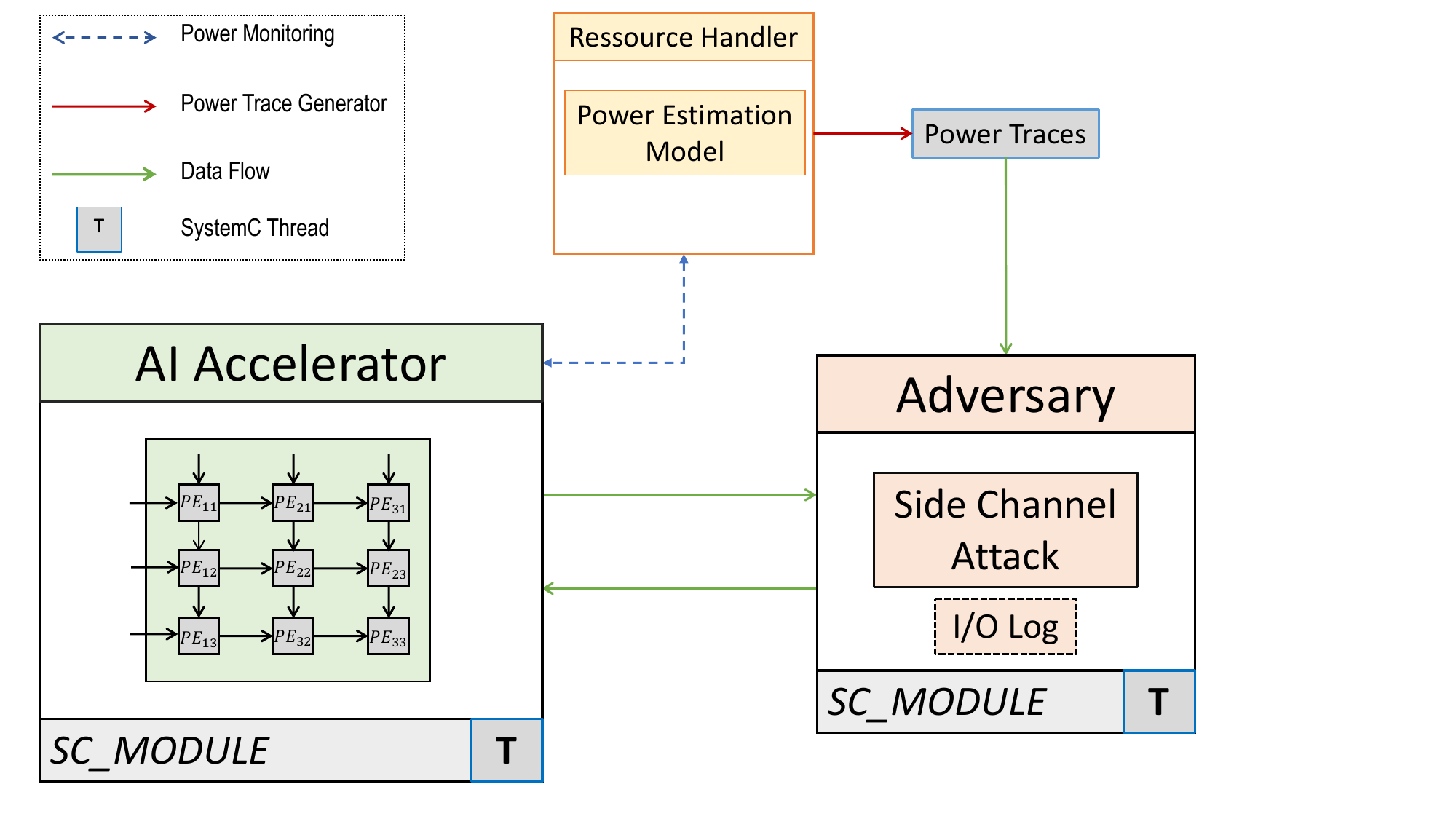}
\centering
\caption{SystemC Implementation Overview.}
\label{fig:RHandler}
\end{figure}

For our approach, we focus on a loosely-timed SystemC model. Here, the behaviour of the AI accelerator is represented by a  SystemC \emph{module} to perform accelerated calculations and mimic the timing and power characteristics of a real hardware accelerator. 
Fig.~\ref{fig:blockDiagram} shows the architecture of the modelled system. The SystemC model easily realizes the individual multiply-accumulate and register operations required during inference, by using dedicated data types.
The matrix multiplication is performed over several cycles depending on the dimension of the matrix by utilizing all \PEs\ in parallel.
The result is therefore available in several parts across multiple cycles as described in Eq.~\ref{eq:PEsInParallel}.
Furthermore, the proposed adversary is implemented as another SystemC \emph{module} which is able to send input and receive output from the AI accelerator.
Lastly, all activity during inference is tracked by a dedicated resource handler shown in Fig.~\ref{fig:RHandler}, which implements a power-estimation model described in the following section.

\section{Dynamic Power Consumption Model}


As the dynamic power consumption (\(P_{Dynamic}\)) is the required power consumption during logical transitions~\cite{JACOB2008847}, the dynamic power-estimation model of a systolic array can be built based on the operations performed by every \PE.
Here, the SystemC model should implement a dedicated resource handler to generate power traces while the calculation is performed~\cite{9129426}.
From ESL perspective, every single operation performed in hardware consumes a certain amount of power measured by the so-called power expense, which depends on the hardware architecture, the type of operations, and the inputs of the operation.
In the following, we utilize the input-dependent power model proposed in~\cite{9129426} and we extend this model to include hardware components such as registers and MAC components.

\subsection{Power Model of a Single Processing Element}
The extended version of the input-dependent model relies on the input of the hardware components and its computational/storage efforts \(CE\).
If the inputs of a hardware component are zero, we consider its contribution to the dynamic power consumption as negligible and its computational/storage effort \(CE\) is zero.
Otherwise, its contribution is not negligible, and its computational/storage effort \(CE\) relies on the number of ones in the input.
The \(CE\) reflects the switching activity of the component and can be described by utilizing a bit-flipping power model.
Several cases of single-bit flipping have to be considered, and a power expense for every case is assigned as follows:
The transition \(0 \rightarrow 0\) and \(1 \rightarrow 1\) require zero power expense, \(0 \rightarrow 1\) requires one power expense, and \(1 \rightarrow 0\) requires 0 power expense as flipping one bit from \(0\) to \(1\) consumes much more power than from \(1\) to \(0\) \cite{9129426}.
The proposed dynamic power consumption model of a single \PE\ estimates the power expense of the MAC component by breaking it down into arithmetic operations.
Additionally, the expense of accessing the register is considered.

\textbf{\textit{MAC Component Power Model:}}
The computational expenses of the MAC component can be broken down into the switching activity of binary arithmetic operations performed during the calculation, namely multiplication and addition.
Counting the flipping of single bits during the calculation provides an estimation of the power expense of the performed MAC operation. Binary arithmetic multiplication can be considered as a series of adders; therefore, the power model of the multiplication is based on the power expenses of a binary adder shown in Table \ref{table:BA}.

\begin{table}[t]
\caption{Power Expenses of Binary Adder}
\resizebox{0.95\hsize}{!}{
\label{table:BA}
\centering
\begin{tabular}{ccccc}
\toprule
\textbf{Input Bits} & \textbf{Output Bits} & \textbf{\( state \) } & \multirow{2}{*}{\textbf{\(CE\)}} &  \multirow{2}{*}{\textbf{\(PM_{BA}\)}} \\
\( (a, b, c) \) & \( (c, s) \) & \textbf{\(expense \)} &  & \\
\midrule
(0,0,0) & (0,0)  & 0  & 0 & 0 \\ 
(1,0,0) & (0,1)  & 0  & 1 & 1 \\ 
(0,1,0) & (0,1)  & 0  & 1 & 1 \\ 
(1,1,0) & (1,0)  & 1 & 2 & 3 \\ 
(0,0,1) & (0,1)  & 1 & 0 & 1 \\ 
(1,0,1) & (1,0)  & 0 & 1 & 1 \\
(0,1,1) & (1,0)  & 0  & 1 & 1 \\ 
(1,1,1) & (1,1)  & 0  & 2 & 2 \\
\bottomrule
\end{tabular}
}
\end{table}

\textbf{\textit{Register Power Model:}}
\(\PM_{Reg}\) denotes the power expenses of the register access power model, which can be modeled based on the bit-switching activity inside the register every time a new value is written.
Therefore, the old and new states of the register (\(Reg_{old}\) and \(Reg_{new}\)) are compared, and the number of switches is counted by using the Hamming Distance (HD):
\begin{equation}
    \PM_{Reg} = \HD(Reg_{old} \oplus Reg_{new}).
    \label{eq:PM_Reg}
\end{equation}

\textbf{\textit{The power consumption model of one PE:}}
The total power consumption of a \PE\ (\(\PM_{\PE}\)) is the sum of the \(\PM_{MAC}\) and the \(\PM_{Reg}\), i.e.,
\begin{equation}
\PM_{\PE} = \PM_{MAC} + \PM_{Reg}.
\label{eq:PM_PE}
\end{equation}

\subsection{Resource Handler}
In the SystemC implementation, the power estimation is performed by the resource handler.
The proposed resource handler relies on the total dynamic power consumption of all \PEs, where
the \PEs\ consume power depending on the performed MAC operation and the register write operation.
These operations are modeled separately and combined to produce a power trace of the whole calculation.
We modify and add our \PE\ power model and utilize the resource handler proposed in~\cite{9129426} to fit the AI accelerator model.
Fig.~\ref{fig:RHandler} illustrates how the resource handler generates power traces of the AI accelerator during inference.

In the following sections, we will show how the power traces generated by the resource handler can be used by an adversary to perform power SCA.
\section{Threat Model}
The proposed threat model is equivalent to a practical one in which the adversary has physical access to an AI edge device~\cite{KotaYOSHIDA20212020CIP0024}.
Regarding the system model, we assume that the adversary has the following capabilities during the attack: 
\begin{itemize}
    \item The adversary has knowledge about the targeted platform or device. 
    \item The adversary has knowledge about the internal structure of the AI accelerator.
    \item The adversary cannot directly access to or read the secret information (weights).
    \item The adversary can input any data into the AI accelerator.
    \item The adversary can observe the device's inference results and obtain power traces of the performed operations.
\end{itemize} 

This scenario can be classified as a grey-box approach~\cite{9149635}, where the target of the adversary is to reveal the \PEs'\ parameters.
These parameters are highly valuable, since they represent the weights of a trained NN. 

Fig.~\ref{fig:blockDiagram} shows an overview of the threat model.
The weight parameters are pre-loaded to the systolic array for inference.
The information leak is caused by the power trace of the inference calculations, thus the adversary can attack the weights via SCA.
\section{SCA Simulation using SystemC}
The described model of an AI accelerator extended with power estimation capabilities enables the modelling of SCA.
Having defined the threat model, we can simulate side-channel attacks targeting the secret parameters of a trained neural network at the ESL.
In order to simulate realistic scenarios, the CPA approach is considered as this approach has been proven successful on real hardware~\cite{KotaYOSHIDA20212020CIP0024}.
In addition, we revisit Template attacks, which are considered the most objective method to assess the leakage of a device under test~\cite{DBLP:conf/eurocrypt/DurvauxSV14,DBLP:conf/crypto/BronchainHMOS19}. 
\subsection{Adversary Simulation in SystemC}
The power estimation model of every \PE\ is a combination of the power estimation models of the single operations performed by the \PE.
Since static power consumption of the device is of no interest for the above mentioned SCAs, the focus lies on dynamic power consumption.
The adversary has access to the modeled power trace and thus can perform the attacks as if the hardware was real. 

\subsection{Correlation Power Analysis}
CPA-based attacks have been proven successful against hardware cryptographic functions~\cite{CPA_general}.
Compared to less complex power analysis attacks, like SPA or DPA, CPA shows a more robust behavior.
To perform a CPA, a leakage model needs to be defined.
The most common approach is to calculate the correlation coefficient between power trace and Hamming Distance (HD) or Hamming Weight (HW) estimation of a certain calculation performed by the observed system.
\begin{figure}[t]
\includegraphics[width=1\linewidth, trim={0.65cm 0.8cm 1.5cm 1.25cm}]{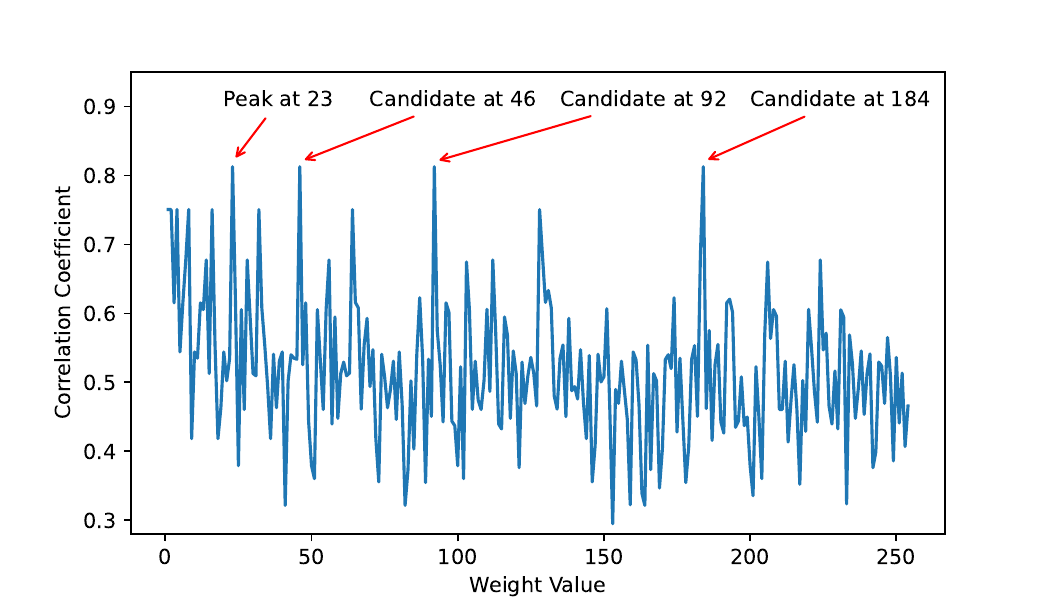}
\centering
\caption{CPA provides multiple weight candidates for \(b_{11}\).}
\label{fig:b11_noNoise}
\end{figure}

Every \(\PE_{ij}\) of the systolic performs a multiply-accumulate operation and stores the result of the operation into a register.
An adversary assumes a correlation between the power traces and the HD model of \PEs\ registers.
To reveal a secret parameter, the adversary calculates the HD estimation ($\hat{H}_{n, b_{k}}$) for all possible transitions of the \(Reg_{ij}\) register by
\begin{equation}
\label{eq:HD_estimation}
    \hat{H}_{n, b_{k}}=\HD\left(Reg_{ij}^{t}, Reg_{ij}^{t+1}\right).
\end{equation}
The correlation coefficient ($\rho\left(b_{k}\right)$) of all estimations and the recorded power traces is calculated as follows: 
\begin{equation}
\label{eq:correlation_coefficient}
\rho\left(b_{k}\right)=\frac{\sum_{n=0}^{N-1}\left(P_{n}-\bar{P}\right)\left(\hat{H}_{n, b_{k}}- \bar{H}_{b_{k}}\right)}{\sqrt{\sum_{n=0}^{N-1}\left(P_{n}-\bar{P}\right)^{2}} \sqrt{\Sigma_{n=0}^{N-1}\left( \hat{H}_{n, b_{k}}-\bar{H}_{n, b_{k}}\right)^{2}}},
\end{equation}
where \(P_{n}\) and  \( \bar{P}\) are the power trace and its average value and $\hat{H}_{n, b_{k}}$ and $\bar{H}_{b_{k}}$ are the HD estimation and its average value. 
The true value of the parameter produces the highest correlation; thus the adversary can reveal it by comparing all of the correlation coefficients as follows: 
\begin{equation}
\label{eq:b_as_argmax}
\hat{b}=\arg \max _{b_{k}}\left(\left|\rho\left(b_{k}\right)\right|\right).
\end{equation}
Since the HD model is not unique for all possible transitions, multiple candidates can provide a similar correlation coefficient (values with bit-shift difference from the true value, e.g. 23, 46, 92, etc.).
This causes certain constraints when revealing the parameters since the attack produces multiple candidates as shown in Fig.~\ref{fig:b11_noNoise}. Nevertheless, the attack can reduce the search space drastically.

\subsection{Template Attack}
Template attacks are a very powerful type of side-channel analysis~\cite{TemplateAttacks}.
As a subset of profiling attacks, template attacks are composed of two phases: profiling and attack phase.
In the profiling phase, the adversary profiles data-dependent power consumption and noise behavior of a target device handling sensitive data.
Then, the adversary performs the attack in the attacking phase to reveal the sensitive data based on the prior knowledge of the device profile. 

In the profiling phase of a template attack, the adversary has full control over a target device and can, e.g., arbitrarily set the weights.
This scenario can be easily simulated with our implemented SystemC model.

Having created the templates for individual \PEs, the adversary can launch the attack by iterating over individual \PEs.
In this attack a small number (10-20) of traces with unknown, but fixed weights leads to a successful recovery.

As the parameters for building the template differ from \PE\ to \PE, the adversary cannot re-use the same template to reveal all of the parameters.
Nevertheless, the additional effort to reveal all of the parameters is only linked to building the template for each of the \PEs.
The acquired traces can be re-used, thus the adversary doesn't require additional power traces (neither for the profiling, nor for the attack).

\section{Attack Results and Impact of Additive Noise}

\begin{figure}[t]
\includegraphics[width=0.8\hsize,trim={2.5cm 0.5cm 3cm 1.3cm}]{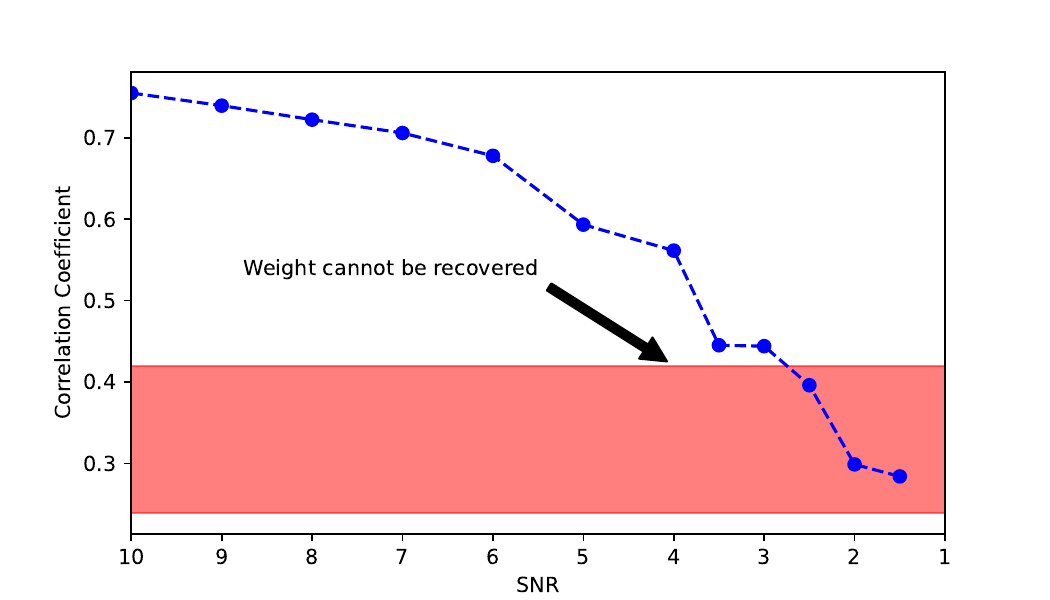}
\centering
\caption{Correlation coefficient against additive noise.}
\label{fig:CorrelationBitByBit}
\end{figure}

Since the model does not consider any measurement noise, the attacker is able to reveal all hidden parameters of the systolic array with the CPA. 
Fig.~\ref{fig:b11_noNoise} shows the simulation results of the attack against the first parameter.
The multiple peaks observed are caused by bit-shifted true values. Since the HD of bit-shifted values is the same, these weight candidates cause very similar correlation levels.
The CPA of a real computing platform is most certainly influenced by measurement noise, therefore, this results should be considered as the best-case scenario (from the attacker's perspective).

The template attack successfully recovers all nine weights from the processing elements with a very low number of attack traces (less than 15 attack traces).
Since we assume an adversary in a chosen-plaintext scenario, the attacker can freely decide which inputs are sent to the systolic array.
By setting entire input columns to zero, the impact of most processing elements which store the pre-loaded weights is eliminated.
This allows the attacker to selectively enable only a small subset (i.e. single columns) of processing elements.
Just like with the CPA, the bit-shifted values of the correct weight produce a high score.
Therefore, it is possible to have multiple candidates as a result of the attack in the leftmost column.
After recovering the weights from the leftmost columns, the attack can build templates including the recovered weights.
This reduces the uncertainty when attacking the middle or rightmost \PEs, thus bit-shifted values of the correct weights do not produce a false positive.

\begin{table}[t]
\caption{Impact of additive noise on attacks}
\label{tab:NoiseImpact}
\centering
\scalebox{0.8}{
\begin{tabular}{ccccc}
\toprule
\multirow{2}{*}{\textbf{Revealed Parameters}} & \multicolumn{2}{c}{\textbf{Template Attack}} & \multicolumn{2}{c}{\textbf{Correlation Power Attack}} \\
\cmidrule(lr){2-3} \cmidrule(lr){4-5} 
 & \multicolumn{1}{c}{\textbf{SNR}} & \multicolumn{1}{l}{\textbf{\# Attack Traces}} & \multicolumn{1}{c}{\textbf{SNR}} & \textbf{Correlation Coefficient} \\
\midrule
9/9 & \multicolumn{1}{c}{$\geq$2.0} & 15 & \multicolumn{1}{c}{\textgreater 4.0}   & 0.561 - 0.775 \\
8/9 & \multicolumn{1}{c}{-} & - & \multicolumn{1}{c}{3.5 - 4.0}          & 0.444 - 0.561 \\
0/9 & \multicolumn{1}{c}{\textless 2.0} & - & \multicolumn{1}{c}{\textless 3.5} & \textless 0.444 \\
\bottomrule
\end{tabular}}
\end{table}

\subsection{Impact of Additive Noise on CPA}
\label{sec:cpa-additive-noise}
In applied cryptography, analysing the impact of additive noise on power attacks is essential~\cite{lerm17cons,st09uni}.
The backbone of such an analysis is Signal-to-noise ratio (SNR) of the leaked information~\cite{CPA-39555-5_29}.
Here, additive noise is used. The measurement noise is modelled by adding random values \(R_{n}\) with an average \(\bar{R}=0\) to the power trace \(P_{n}\) at each estimation point as  \(P_{n}+R_{n}\).
It can be gradually increased to have a bigger impact on the power estimation value.
Here, we can set a fixed SNR to produce noisy power traces.

With this model, a threshold evaluation of the CPA's success is possible. 
Multiple experiments with additive noise are performed to investigate the influence of noise on correlation coefficients.
A comparison of the correlation with different amounts of additive noise is shown in Table~\ref{tab:NoiseImpact} and illustrated in Fig.~\ref{fig:CorrelationBitByBit}.
The results show how an increasing noise level impacts the correlation coefficient, ultimately making the correct candidate indistinguishable from other candidates.
For too low SNRs, the CPA cannot successfully reveal the weights from the AI accelerator. Increasing the number of traces an attacker acquires, increases the chance of a successful attack. This can give an indication to how many traces an attacker would require in a post-silicone attack.

\subsection{Impact of Additive Noise on Template Attacks}
Several experiments have been conducted to study the impact of additive noise on template attacks, where both the profiling and attack traces are affected. 
SNR is also used to describe the magnitude of the noise. The experiments show that recovering the weights remains as easy as without noise.
By increasing the impact of noise, i.e., decreasing the SNR to as low as 2.0, the template attacks proves to be successful with as little as 15 attack traces per targeted parameter.

Consequently, template attacks are applicable with lower SNR values, i.e., the template attacks are much less affected by noise if the same noise level is present during the profiling phase, as well as the attack phase.
Template attacks therefore, pose a serious threat to implementations, even in a noisy environment.
By taking advantage of input tuning (as a chosen plaintext attack), an adversary could theoretically attack systolic arrays of any size and reveal secret parameters. Here, a more noisy environment requires the attacker to use a larger number of traces when building the template.
\section{Model Verification}

\begin{figure}[t]
\includegraphics[width=\hsize, trim={0.5cm 1cm 4cm 0.5cm}]{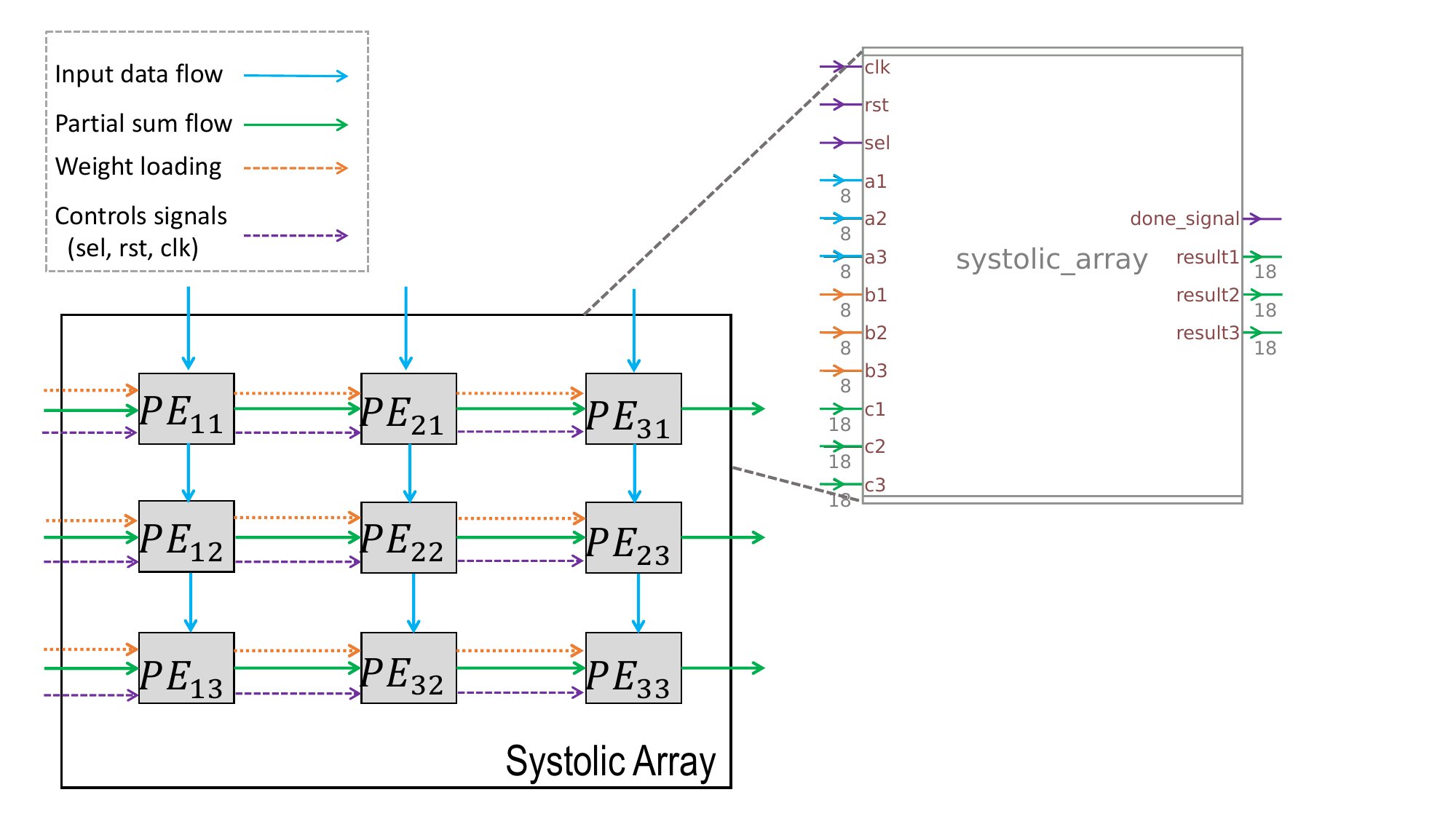}
\centering
\caption{Systolic Array Implemented for Verification.}
\label{fig:Diagram:SystolicArray}
\end{figure}

It was previously shown \cite{SynopsysPP}, that a time-based power estimation with a gate-level netlist comes quite close to post-silicon measurements. 
To achieve industry-grade results, we also use the Synopsys tool suite for our experiments.

The verification starts with a Verilog implementation of a single PE, as well as the whole \((3 \times 3)\) systolic array, as shown in~\ref{fig:Diagram:SystolicArray}. The design is synthesized using Synopsys DesignCompiler \cite{SynopsysDC} to generate a gate-level netlist. A pre-defined test bench is used to stimulate the netlist and gather a value-change dump (VCD) using Synopsys VCS \cite{SynopsysVCS}. Lastly, Synopsys PrimePower \cite{SynopsysPP} creates a power trace based on the VCD in a time-based power analysis. These power traces are considered to be noise-free reference traces of the real hardware. We use these traces to verify the input-dependent power model utilized in our SystemC simulation of a systolic array. Consequently, a statistical comparison between the reference power traces and the power traces collected at the SystemC level is performed. The comparison is divided into four main experiments as follows: 

\begin{figure}[t]
\includegraphics[width=\hsize]{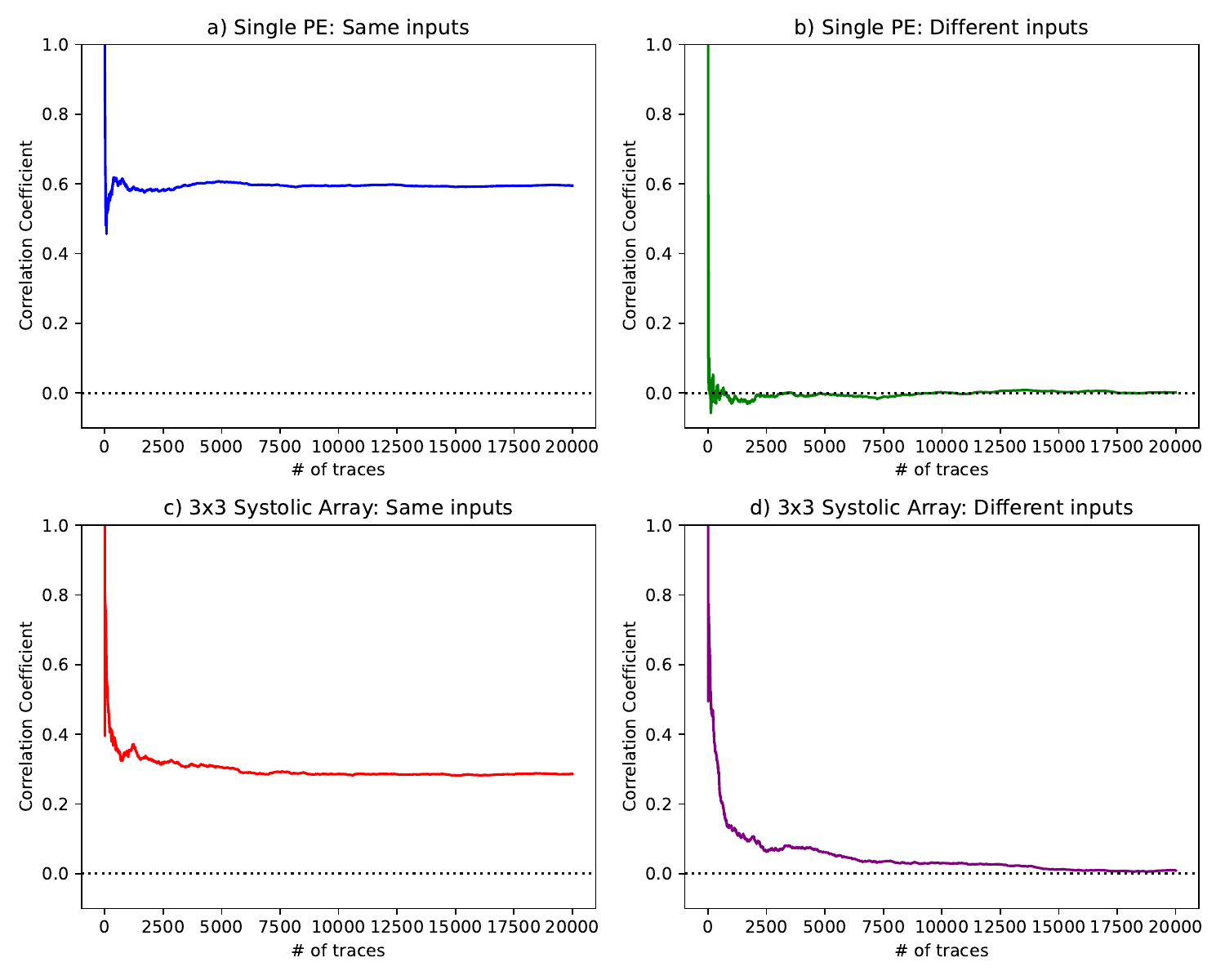}
\centering
\caption{Pearson's Correlation Coefficient between Trace Sets.}
\label{fig:Traces-Correlation}
\end{figure}

\subsection{First Experiment}
For a single PE with the same random inputs, we generate two sets of power traces (20\,000 traces) collected at SystemC level and gate-level netlist. Then, we use Pearson's correlation coefficient (PCC) to interpret if there is a linear correlation between them. PCC lies between \([-1, +1]\), with \(PCC=0\) indicating no linear correlation. The results show that there is a positive correlation between the traces, as seen in Fig.~\ref{fig:Traces-Correlation}a. A value of PCC up to 0.65 confirms that the proposed SystemC model is linearly associated with the power consumption tendency of a real hardware implementation. 
\subsection{Second Experiment}
For a single PE with two \textit{distinct} sets of random inputs, we generate two sets of power traces (20\,000 traces) collected at SystemC level and gate-level netlist. The goal of this experiment is to exclude a false-positive correlation for a single PE. Here, we observe a correlation coefficient close to 0, as shown in Fig.~\ref{fig:Traces-Correlation}b. This confirms there is no false positive correlation between the power traces. 
\subsection{Third Experiment}
The design of the whole systolic array is more complex. A test bench with full coverage of all possible input/weight combinations for all PEs would produce an enormous amount of traces. Therefore, we fixed the weights in the PEs and stimulated the systolic array by 20\,000 random input samples. An equivalent test bench is implemented in SystemC to produce comparable traces.
As we expected, the two trace sets will be linearly correlated the most when modeling smaller pieces of hardware. Naturally, modelling bigger hardware at a high abstraction level will bring a drop in accuracy, and PCC will be lower, as shown in the in Fig.~\ref{fig:Traces-Correlation}c. Therefore, Spearman's correlation coefficient (SCC) is used in this experiment to interpret the direction of the association between them. The sign of the SCC value indicates if the same trends are expected between the two trace sets. When evaluating SCC between the two sets, we observe a positive SCC coefficient of \(+0.27\). This indicates a moderate monotonic (linear or non-linear) relationship between them. In other words, SCC shows that power traces collected at SystemC level tend to increase when reference power traces increase.

\subsection{Fourth Experiment}
Similar to the second experiment, we aim to exclude a false-positive correlation result between the traces of the whole systolic array. With two distinct sets of random inputs, we observe both PCC and SCC close to 0, as shown in Fig.~\ref{fig:Traces-Correlation}d. This indicates there is no false positive in the power traces collected at SystemC level.

In conclusion, it can be said with high confidence that the proposed power estimation model follows the same trends as a state-of-the-art netlist-level power estimation.
\section{Conclusion}
This paper presents power side-channel attacks against AI accelerator architectures at the electronic system level. Our approach features AI accelerator models with a corresponding dynamic power-consumption model to simulate the behaviour of systolic-array-based AI accelerators using SystemC. Our findings show that SystemC-based power attacks are possible and sufficiently resemble real-world threat scenarios. Our experiments successfully simulate SystemC-based power side-channel attacks against AI accelerators leading to full secret extraction:
While correlation power analysis shows certain limitations in noisy conditions, template attacks pose a significant risk of being able to adapt to noise. 

To verify the SystemC-power estimation model, several experiments were performed to compare power traces computed from synthesized netlists with the proposed model. The results show that the proposed model follows the same trends as a gate-level netlist power estimation. Our set goal of earliest-possible threat analysis and subsequent design suggestions was thus successfully achieved and demonstrated.

This work hence is one essential -- and with regard to the presented methods and procedures to the best of our knowledge first -- step in design-space exploration for security from a design/hardware perspective. In a future step and raising complexity, we would like to extend this approach from a systolic array to a full system model.

\bibliographystyle{IEEEtran}
\bibliography{sample-base}
\vspace{12pt}

\end{document}